\documentclass{article}

\usepackage{PRIMEarxiv}

\usepackage[utf8]{inputenc} 
\usepackage[T1]{fontenc}    
\usepackage{hyperref}       
\usepackage{url}            
\usepackage{booktabs}       
\usepackage{amsfonts}       
\usepackage{nicefrac}       
\usepackage{microtype}      
\usepackage{lipsum}
\usepackage{fancyhdr}       
\usepackage{graphicx}       
\graphicspath{{media/}}     
\usepackage{subcaption}
\usepackage{graphicx}
\usepackage{multirow}
\usepackage{multicol}

\title{Using LoRa communication for Urban VANETs: Feasibility and Challenges
}

\author{
  Thenuka Karunathilake \\
  University of Bremen \\
  Bremen, Germany\\
  \texttt{thenuka@comnets.uni-bremen.de} \\
   \And
  Anna F\"orster \\
  University of Bremen \\
  Bremen, Germany\\
  \texttt{anna.foerster@uni-bremen.de} \\
}

\begin{document}
\maketitle

\begin{abstract}
Vehicular Ad-Hoc Networks (VANETs) were introduced mainly to increase vehicular safety by enabling communication between vehicles and infrastructure to improve overall awareness. The vehicles in a VANET are expected to exchange numerous messages generated by multiple applications, but mainly, these applications can be subdivided into safety and non-safety. The main communication technologies designed for VANETs, DSRC (Dedicated Short Range Communication) and C-V2X (Cellular V2X), mainly focus on delay-sensitive safety-related applications. However, sharing the same bandwidth for safety and non-safety applications will increase the burden on the communication channel and can cause an increase in the overall latencies. Therefore, this work analyses the feasibility of using LoRa communication for non-safety-related urban VANET applications. We conducted multiple real-world experiments to analyse the performance of LoRa communication in various urban VANET scenarios. Our results show that LoRa communication handles the Dopper shifts caused by the urban VANET speeds with both Spreading Factor (SF) 7 and 12. However, higher SF was more vulnerable to Doppler shifts than lower SF. Furthermore, the results illustrate that the Line-of-Sight (LoS) condition significantly affects the LoRa communication, especially in the case of lower SF.
\end{abstract}

\keywords{LoRa \and Vehicular Networks \and Smart Cities}

\section{Introduction}
 \label{sec:Section 1}
  
In recent years, the applications of Vehicular Ad-Hoc Networks (VANETs) have increased significantly, and VANETs have become the foundation for Intelligent Transportation Systems (ITS) in future smart cities. There are many applications of VANETs, mainly safety message exchange, platooning, road traffic control, service advertisements, parking spot identification, and infotainment. In many ways, VANETs are similar to conventional Wireless Sensor Networks (WSNs), but some major significant differences exist due to the mobility of the nodes. In WSNs, most nodes can be considered static or moving at relatively low speeds, but in contrast, nodes associated with VANETs have higher moving speeds. Therefore, traditional communication technologies do not work reliably with VANETs due to frequency shifts caused by the Doppler effect. Hence, several communication technologies are specifically designed to work with VANETs and handle their special characteristics. 

The main communication technologies associated with VANETs are Dedicated Short Range communication (DSRC) \cite{kenney_dedicated_2011} and Cellular V2X (C-V2X) \cite{khan_advancing_2023}. DSRC was the initial technology specially designed for VANET applications. It is based on IEEE 802.11p (an improved version of 802.11 to handle mobility) and can handle both Vehicle-to-Vehicle (V2V) and Vehicle-to-Infrastructure (V2I) communications. After its introduction, DSRC was heavily tested in many VANET applications, and now it has become the most mature VANET communication technology. However, DSRC has several drawbacks: relatively expensive equipment cost, shorter range, and carrier sense multiple access (CSMA) mechanism. With CSMA, all network nodes must check the channel activity before transmitting data. The CSMA mechanism can guarantee low latency in low-dense networks due to fewer competing nodes. However, in a high-dense network, the CSMA mechanism can increase latency. 

Recently, C-V2X technology, which is based on cellular technologies like LTE and 5G, was introduced to VANETs. C-V2X can handle longer distance communications than DSRC with network-based communication through base stations, additionally to direct V2X communications. Furthermore, C-V2X has the ability to communicate with pedestrians through cellular-operated mobile devices which is not possible with DSRC technology. However, C-V2X suffers from longer latencies due to a semi-persistent scheduling mechanism where nodes select random resource blocks for transmission. 

The current market penetration of the number of connected vehicles (V2X capable vehicles) is relatively small, but it is expected to increase significantly in the near future. In a highly congested traffic scenario (ex, intersections and crowded highway sections), the available bandwidth of DSRC and C-V2X technologies may not be enough to cater to all the connected vehicles simultaneously. In our previous work \cite{karunathilake_survey_2022}, we calculated how many connected vehicles can be served by a single 5G base station, and the results illustrated that in highly congested scenarios, the available bandwidth is not sufficient, which can lead to increased latency and hinder the vehicular safety. Therefore, it is beneficial to incorporate another supporting communication technology for non-safety related applications to ease the burden on the main communication technologies (DSRC and C-V2X) for such congested scenarios, allowing primary technologies to focus on safety-related applications solely. 

Usually, most congested scenarios occur in urban areas due to the high density of vehicles. Therefore, this work extensively investigates the feasibility and challenges of using LoRa communication as a supporting communication technology in urban VANETs. Since LoRa communication is designed for WSN devices, the performance of LoRa has to be investigated in challenging urban VANET characteristics like the presence of Non-LoS communication and Doppler shifts. Therefore, we conducted multiple real-world experiments in different locations with different speeds in V2V and V2I configurations. The performance of LoRa is analysed in terms of packet delivery ratio and Received Signal Strength Indicator (RSSI) with different LoRa SFs in urban speeds up to 58 km/h. 

The rest of the paper is arranged as follows. In Section \ref{sec:Section 2}, we describe the details of LoRa communication, and in Section \ref{sec:Realted}, we discuss prior works done on LoRa communication in V2X scenarios. The used experiment setup and scenarios are explained in detail in Section \ref{sec:Experiment Setup}, and in Section \ref{sec:Section 4}, a detailed analysis of the results is provided. Finally, in Section \ref{sec:Section 5}, we conclude our work. 

\section{LoRa Communication}
\label{sec:Section 2}

Long Range (LoRa) communication has gained extensive attention after its introduction in 2015, mainly in the domain of Wireless Sensor Networks (WSN) and IoT. LoRa has several key benefits over other existing communication technologies, mainly the extended communication range (10 kilometers in optimal condition), extremely low power consumption, etc. These characteristics have made LoRa an ideal solution for WSNs. However, LoRa also has some disadvantages, mainly the relatively low data rate (0.3 kbit/s to 50kbit/s per channel) \cite{idris_survey_2022}, which limits the use of LoRa communication in applications where high-speed data links are essential.

LoRa is a mostly used LPWAN technology in WSN and IoT applications \cite{idris_survey_2022}. It is based on the Chirp Spread Spectrum (CSS) modulation technique, which uses chirp signals to transmit information. Over time, chirp signals can increase in frequency (up-chirp) or decrease (down-chirp). LoRa uses sub-GHZ unlicensed ISM frequency bands for communication, and frequency bands depend on the country or region (in Europe, 433MHz and 863MHz). There is a common confusion between LoRaWAN and LoRa in the literature, and they are often used synonymously. However, there is a significant difference between these two terms. LoRa handles the physical layer of the protocol stack with CSS modulation for long-range communication. In contrast, LoRaWAN handles the MAC layer, enabling communication between LoRa gateways and duty-cycled end devices (class A,B and C). LoRaWAN specifications are developed and maintained by the LoRa alliance, which consists of multiple companies and research bodies.

In our real-world experiments, we only use the LoRa physical layer to allow direct communication between LoRa nodes without LoRaWAN. 

\section{Related Works}
\label{sec:Realted}

In this section, we discuss some prior work done to evaluate LoRa communication in the context of VANETs.

As discussed earlier, the Doppler shift is a significant concern when using LoRa in highly mobile scenarios. Therefore, authors of \cite{li_lora_2018} performed simulation analysis to determine which LoRa parameters should be selected in fast-fading scenarios caused by the Doppler effect. The results show that higher bandwidth and lower spreading factors are more suitable for V2X scenarios.

One of the major benefits of LoRa communication is its long communication range. However, these longer communication ranges depend on many factors, including LoS communication, mobility, and terrain distribution. Therefore, the authors of \cite{oliveira_lora_2021} conducted an experiment in an urban scenario with one moving transmitter and a static gateway. The results show that the obtained delivery ratios depend significantly on speed. Furthermore, the collected RSSI and SNR depend on terrain distribution and LoS communication rather than the distance between the transmitter and the gateway. A similar work was carried out by the authors of \cite{sanchez-iborra_performance_2018} and \cite{alvear_assessing_2018}. In \cite{sanchez-iborra_performance_2018}, experiments were conducted in multiple environments, including urban, suburban, and rural environments where the average moving speed was different. The results illustrate that the LoRa communication range is significantly longer in rural scenarios compared to urban scenarios. Moreover, in mobility scenarios, LoRa performance depicted a vulnerability related to the Doppler effect.  

VANETs consist mainly of V2I and V2V communications, and in V2V scenarios, the relative speed varies more than in V2I scenarios. The authors of \cite{torres_experimental_2021} conducted experiments in a University environment for both V2I and V2V scenarios independently. Initially, 50 different locations were selected, and later, it was reduced to 9 locations by simulation analysis. Two vehicles were used for the experiments, and in the V2I scenario, one vehicle was stopped at these selected locations while the other vehicle was moving around the university. In the V2V scenario, both vehicles were moved randomly. The results show that higher spreading factors have lower RSSI values in mobility scenarios, and lower spreading factors depict lower communication ranges.

A LoRa-based architecture for V2X communication was proposed in \cite{haque_lora_2020}. The authors used multiple RSUs and a vehicle equipped with an On-Board-Unit (OBU) for the experiments. In the proposed architecture V2V and V2I communication decisions are taken based on the RSSI values. In the case of V2I, when multiple RSUs are in the communication range, the vehicle decides to communicate with the RSU, which has the best channel quality (RSSI). Once the RSSI value decreases below a preset threshold value the vehicle handover the communication to another RSU which has better channel quality. The results show that in the V2I configuration, the proposed architecture performed as expected, up to the speed of 50 km/h.

All the above-mentioned prior work evaluates LoRa communication performance in mobility scenarios. However, these evaluated scenarios combine LoS communication, Non-LoS communication, and different Doppler shifts. Therefore, each of these characteristics cannot be investigated independently. As a solution, we conducted experiments in multiple locations to isolate these characteristics. For example, in our LoS experiments, LoS communication is guaranteed during the whole experiment, and we kept the moving speed of the transmitter and receiver at a constant value during the experiments. Hence, each of the characteristics can be isolated and can be analysed independently.

\section{Experiment Setup}
\label{sec:Experiment Setup}

This section describes our hardware setup and used scenarios for the experiments in detail. 
 
\begin{figure}[]
\centering
\includegraphics[width=0.5\textwidth]{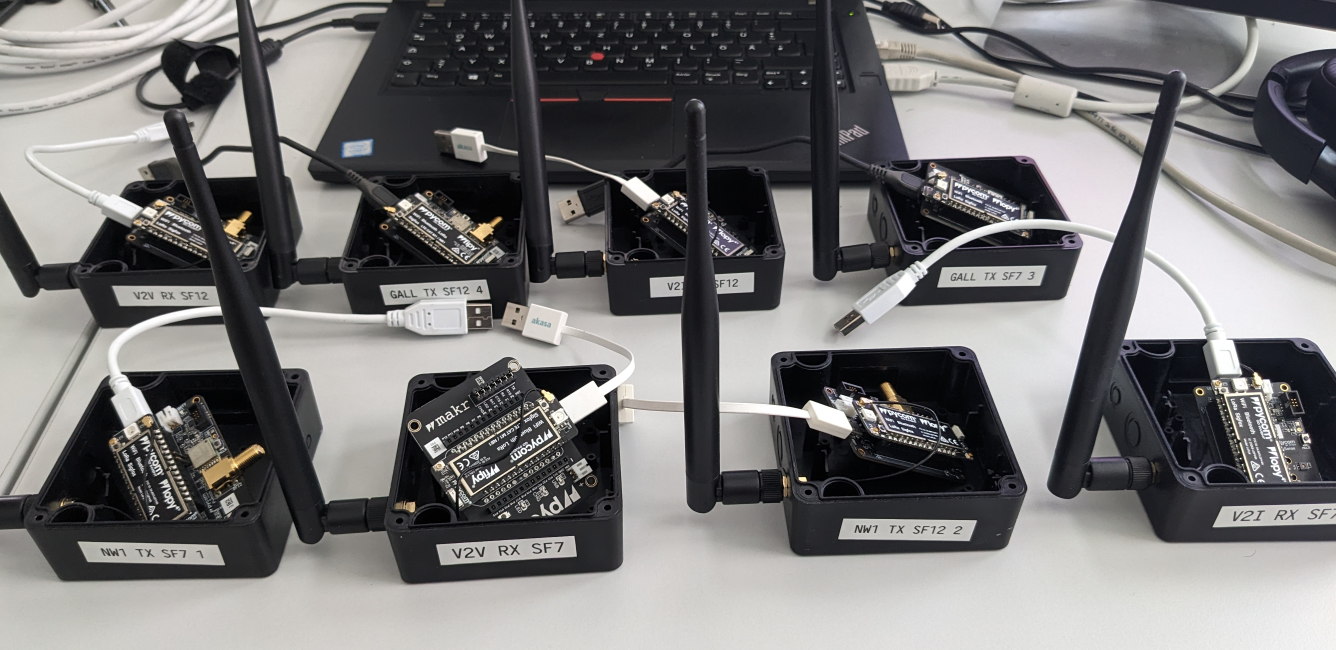}
\caption{Experiment Setup}
\label{Exp_Setup}
\end{figure}

We used multiple Pycom LoPy4 devices to conduct the experiments, as shown in Figure \ref{Exp_Setup}. Pytrack expansion boards were used to extract GPS coordinates with LoPy4 as the transmitters. For the receivers, a combination of expansion boards, pysense, and pytracks was used with the LoPy4 devices. All the LoPy4 devices were configured with default LoRa parameters as shown in Table \ref{parameters}, two different spreading factors were used (SF-7 and SF-12), and the packet transmission interval was set to 1 second.

We mainly divide our experiments into two scenarios. The first set of scenarios was conducted with LoS communication, therefore named as 'LoS Scenario' in the following. Later, we performed another set of experiments in a scenario with both LoS and Non-LoS communication because, in real-world urban VANET scenarios, both LoS and Non-LoS are combined. This scenario is named as 'LoS + Non-LoS Scenario' in the following.

\begin{table}
\centering
\footnotesize
\caption{LoRa Parameters for all our experiments}
\label{parameters}
\begin{tabular}{|c|c|} \hline
Parameter &  Values \\ \hline
Mode & Raw lora \\ \hline
Region  & EU 868 \\ \hline
Frequency  & 868 MHz\\ \hline
TX power & 14 dBm \\ \hline
Bandwidth  & 125 kHz \\ \hline
Spreading Factor (SF)  & 7 and 12\\ \hline
Packet Interval  & 1 s\\ \hline
\end{tabular}
\end{table}

\subsection{LoS Scenario}
\label{sec:LOS Scenario}

\begin{figure}[]
\centering
\includegraphics[width=0.5\textwidth]{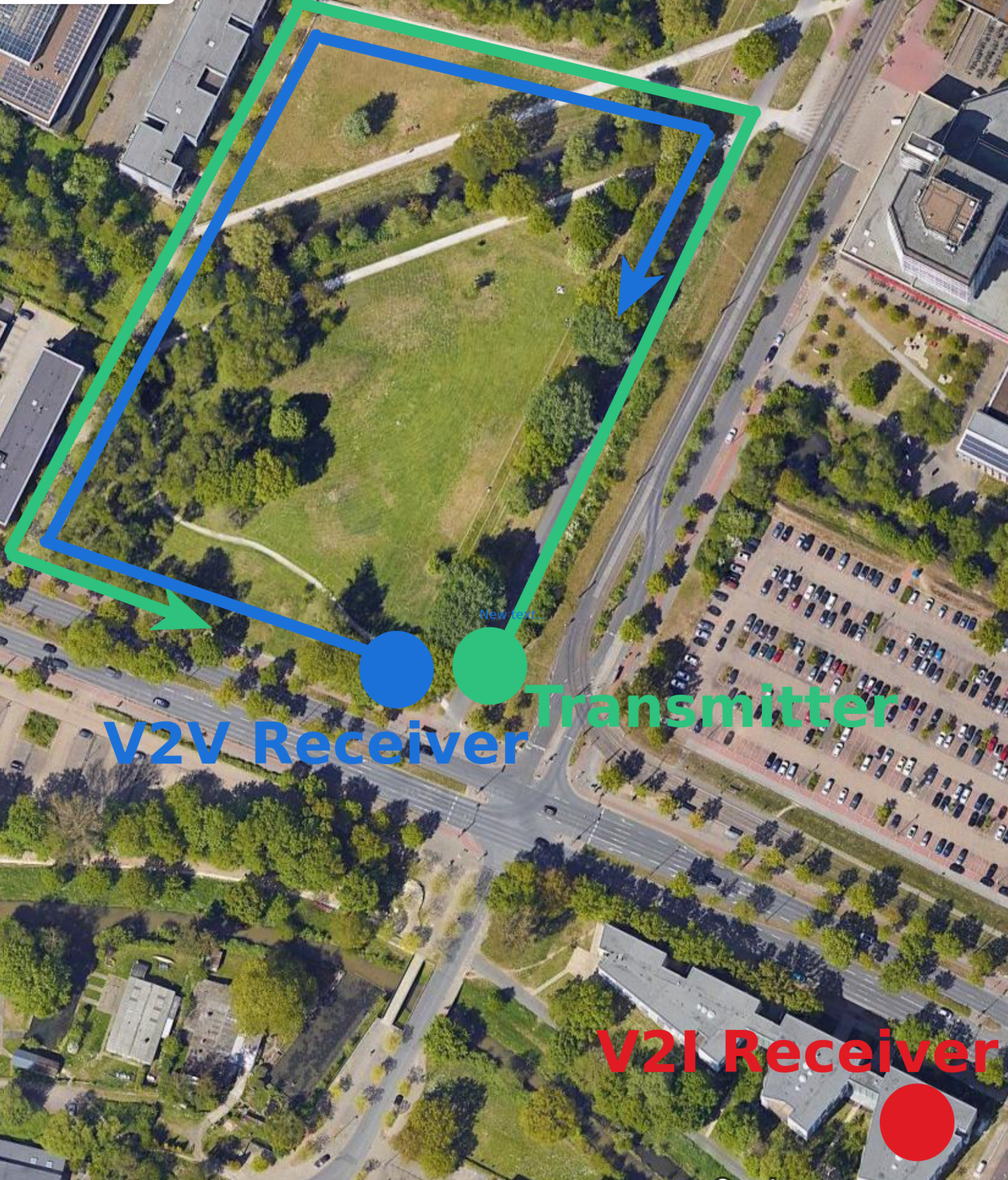}
\caption{LoS experiment scenario}
\label{los}
\end{figure}

The campus park (Figure \ref{los}) of the University of Bremen was used for LoS experiments, and the route around the campus park is approximately 650 m long. Three devices (one transmitter and two receivers) were used during the experiments. The transmitter and one of the receivers were moved on this route in opposite directions (to have the maximum relative speed), as shown in Figure \ref{los}. Since this receiver is moving, it is considered to be the 'V2V receiver'. The other receiver was placed in a nearby building, as shown in Figure \ref{los} with red color. This receiver is a stationary device; therefore, we consider it to be the 'V2I receiver'. The V2I receiver was placed outside a window of the building to improve LoS communication. In the initial set of experiments, V2I and V2V configurations were conducted as two separate experiments using only 2 devices (one transmitter and V2I receiver or V2V receiver). However, we later conducted V2I and V2V experiments simultaneously using all 3 devices (one transmitter and V2I receiver and V2V receiver). 

We conducted multiple experiments by changing the moving speed of the transmitter and the V2V receiver. Each experiment was conducted for 30 minutes, and we conducted experiments with different speeds of 4 km/h, 10 km/h, and 20 km/h. Both the transmitter and V2V receiver were moved at the same speed during one experiment. For example, in the case of 20 km/h, both the transmitter and V2V receiver were moved at 20 km/h for 30 minutes, and during the entire 30 minutes, we tried to maintain the same constant speed. However, for safety reasons, we had to reduce the speed slightly during cornering maneuvers at higher speeds (10 km/h and 20 km/h). Furthermore, in the V2V configuration, the maximum relative speed can reach up to twice the moving speed of one node (20 km/h if the transmitter and receiver are moving at 20 km/h). Different modes of transportation were used to achieve different speeds. The experiments with 4 km/h were conducted by walking around the route with two persons. One carried the transmitter, and the other person carried the V2V Receiver. Similarly, 10 km/h was achieved by using bicycles, and 20 km/h was achieved by using electric scooters. We used speed-measuring applications on mobile devices to measure the moving speed during the experiments.

The first set of experiments was conducted in the Summer of 2022, and in the following sections, we refer to these experiments as 'LoS-Summer'. Even though no buildings affect the LoS communication, still, as shown in Figure \ref{los}, some trees might slightly affect LoS conditions. Therefore, the same set of experiments was conducted again in the Fall of 2022 because, during the fall season, better LoS communication was expected because there were no leaves on the trees. In the following, we refer to these experiments as 'LoS-Fall'.

The main goal of the work was to check the feasibility of using LoRa communication for urban VANET applications. Since the driving speeds of vehicles can be higher than 20 km/h in urban scenarios, more experiments had to be conducted to investigate the LoRa communication behavior with higher speeds. However, in reality, it is difficult to achieve higher speed while maintaining LoS communication. For example, other road vehicles (cars, motor bicycles and etc) can be used to achieve higher speeds, but in reality, it is not possible to maintain the same constant speed due to many obstacles like traffic lights, pedestrian crossings, etc., while maintaining LoS communication. 

\begin{figure}[]
\centering
\includegraphics[width=0.5\textwidth]{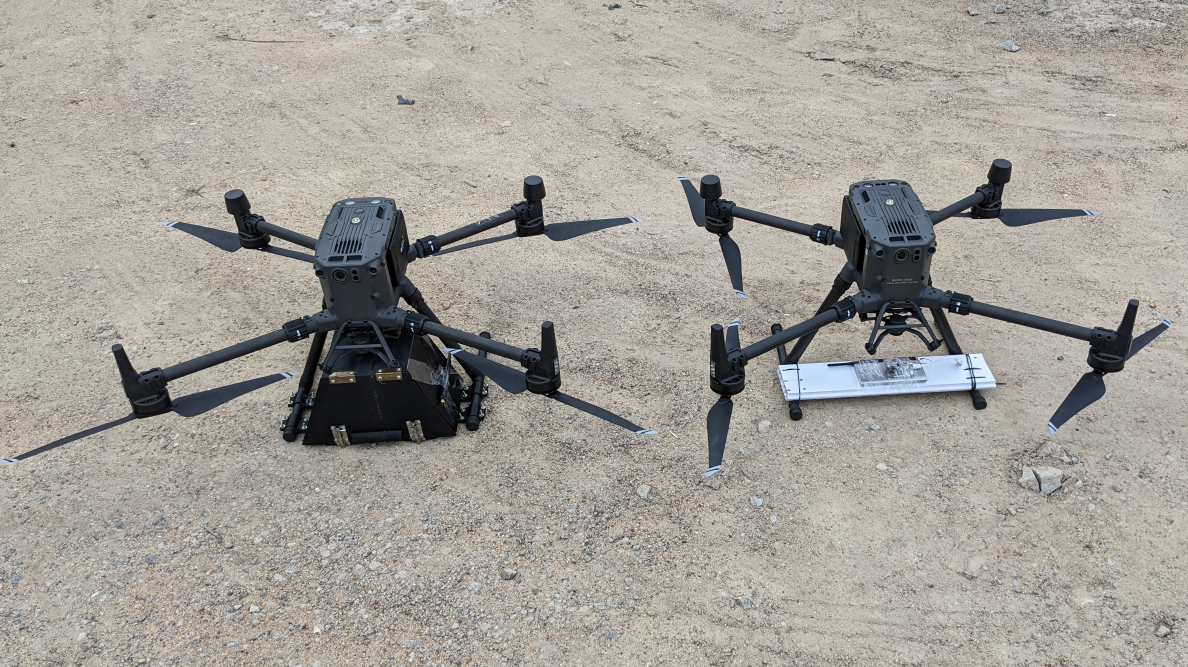}
\caption{The used drones with attached Pycom devices}
\label{drone_pic}
\end{figure}

\begin{figure}[]
\centering
\includegraphics[width=0.5\textwidth]{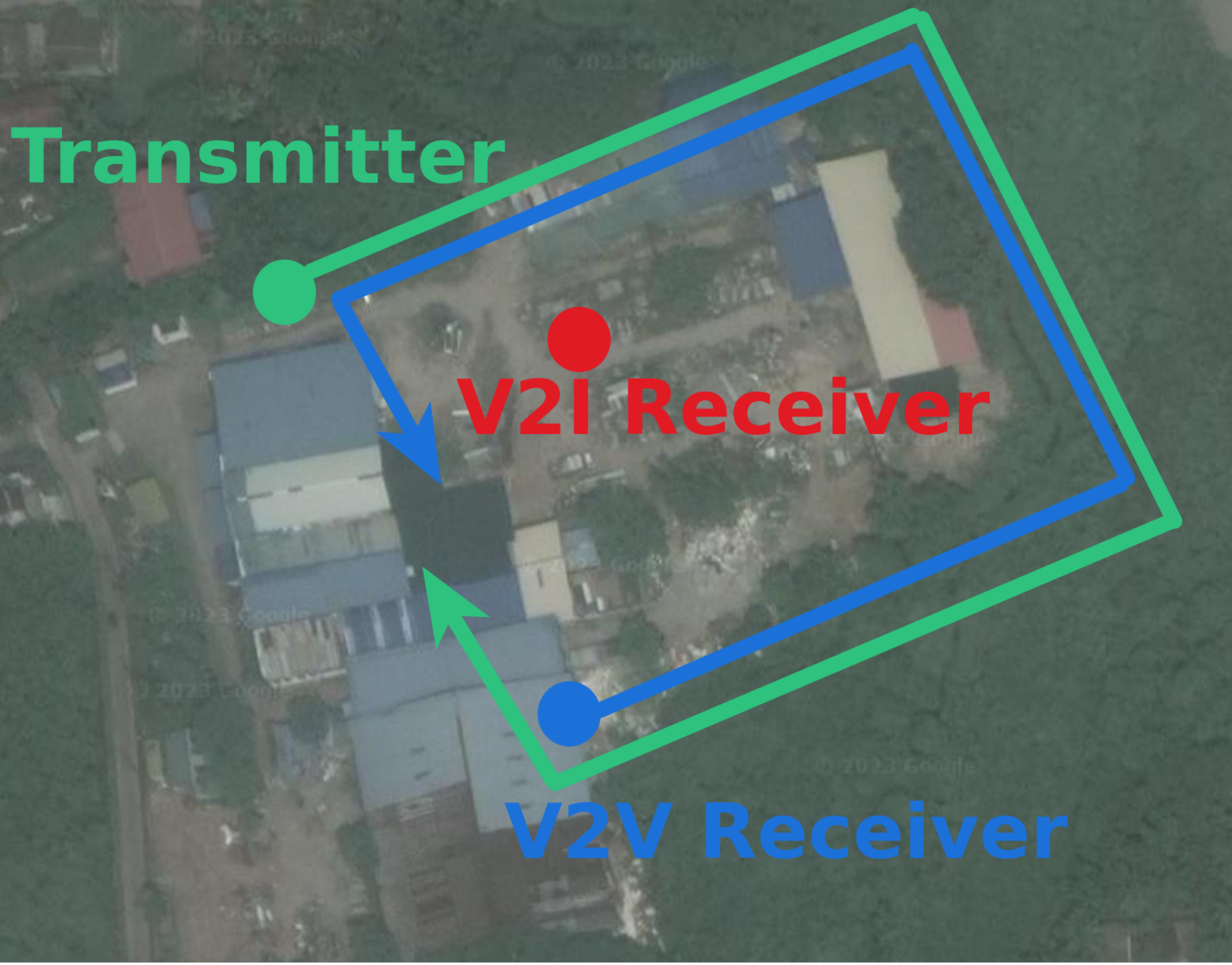}
\caption{LoS experiment scenario with drones}
\label{los-drone}
\end{figure}

To overcome these issues, we decided to use two drones to conduct experiments with higher speeds. Since the drones fly above ground, the LOS communication and constant speed are guaranteed. We decided to use two DJI Maverick M300 RTK drones for the experiments, as shown in Figure \ref{drone_pic}. We selected this specific drone model because it was able to reach the expected speed levels while carrying the Pycom devices and the battery pack. These drones had a maximum speed limit of 58 km/h; therefore, we conducted 40 km/h and 58 km/h experiments. Predefined waypoints were used to define the path for the drones, and the drone was able to maintain a constant speed while counteracting the wind speed. The drone experiments were conducted in an open area without obstacles in Colombo, Sri Lanka, as shown in Figure \ref{los-drone}. One drone was used to carry the transmitter, the other drone was used as the V2V Receiver, and the V2I Receiver was placed on the ground level. The drone with the transmitter was set to fly at a height of 70 m, and the V2V Receiver drone was flown at a height of 50 m to avoid collision between the drones. Furthermore, each drone experiment was limited to 15 minutes due to the battery life of the drones.

\subsection{LoS + Non-LoS Scenario}

\begin{figure}[]
\centering
\includegraphics[width=0.5\textwidth]{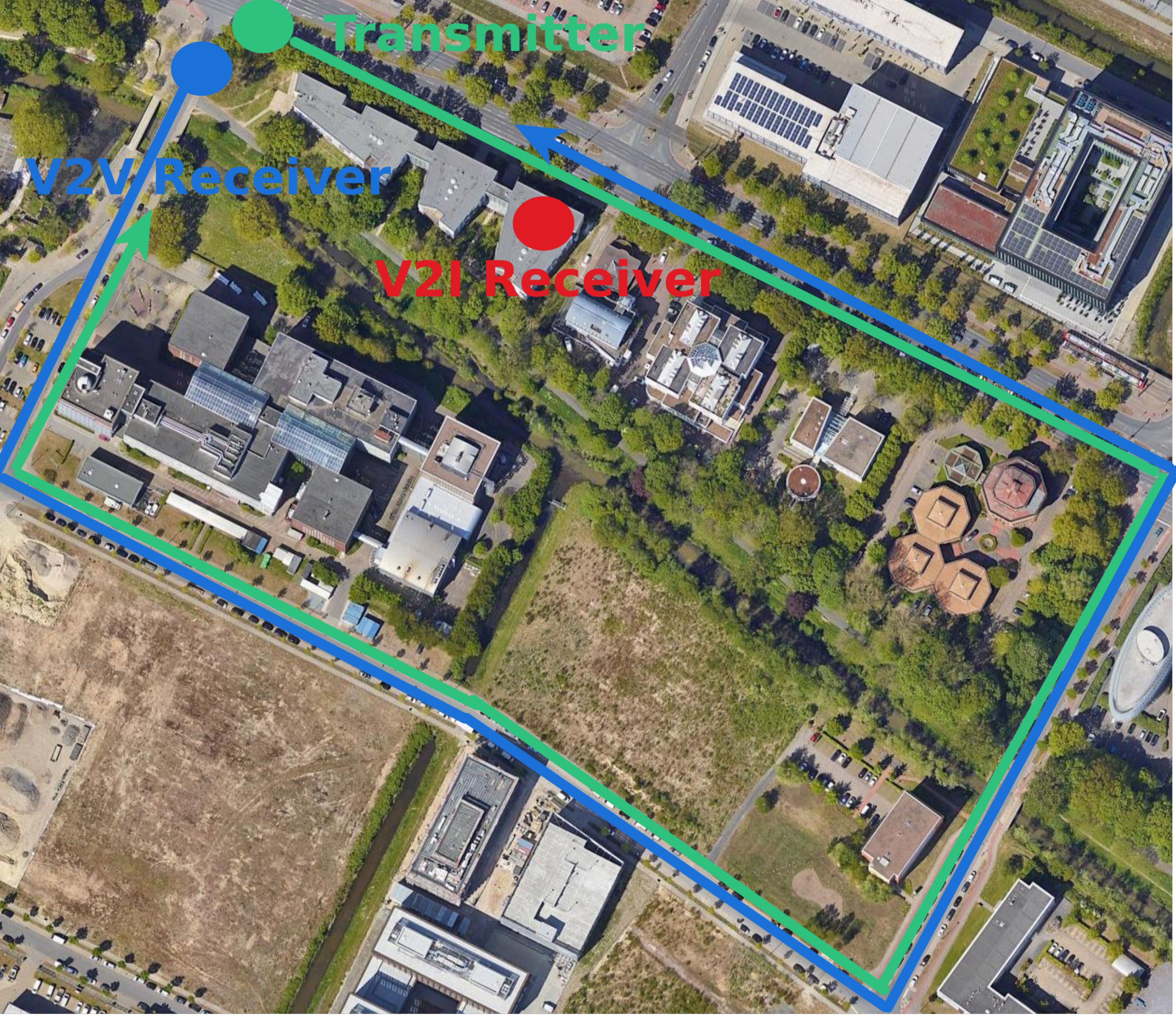}
\caption{LoS + Non-LoS experiment scenario}
\label{non-los}
\end{figure}

In real-world urban VANET scenarios, it is impossible to achieve LoS communication continuously because of other objects (buildings, trees, etc.). Therefore, most VANET scenarios are a combination of both LoS and Non-LoS communication. Hence, we extended our experiments with another scenario by introducing both LoS and Non-LoS communication. The 1.2 km long route around the NW1 building of the University of Bremen, Germany, was selected to conduct LoS + Non-LoS experiments. As shown in Figure \ref{non-los}, multiple buildings are located in the surroundings; therefore, LoS communication is not guaranteed. We conducted a similar set of experiments as explained in the section \ref{sec:LOS Scenario} with LoS + Non-LoS scenario. However, LoS + Non-LoS scenario experiments were only conducted during the Summer of 2022 because there was no significant difference between the Fall and the Summer for this scenario. Furthermore, the LoS + Non-Los Scenario was only investigated with speeds up to 20 km/h.

\section{Results Analysis}
\label{sec:Section 4}

In this section, we analyse the obtained results from the experiments. As explained in Section \ref{sec:Experiment Setup}, we conducted experiments in the Summer and Fall seasons. Therefore, our analysis in this section is categorised into two main parts: Summer experiments and Fall experiments.  

\subsection{Summer experiments}
\label{subsec:Summer-Experiments}

 \begin{table*}
\begin{center}
\resizebox{\textwidth}{!}{
\begin{tabular}{|l|l|l|l|l|l|l|l|l|l|l|l|l|l|l|}
\hline
\multicolumn{1}{|c|}{\multirow{2}{*}{\textbf{Scenario}}}& \multicolumn{1}{|c|}{\multirow{2}{*}{\textbf{Speed}}}&\multicolumn{1}{|c|}{\multirow{3}{*}{\textbf{{\begin{tabular}{@{}c@{}}Mode of Transport \\ and Location\end{tabular}}}} }& \multicolumn{6}{|c|}{\textbf{V2I Configuration}} & \multicolumn{6}{|c|}{\textbf{V2V Configuration}}  \\
\cline{4-15}
& & & \multicolumn{3}{|c|}{\textbf{SF 7}} & \multicolumn{3}{|c|}{\textbf{SF 12}}& \multicolumn{3}{|c|}{\textbf{SF 7}} & \multicolumn{3}{|c|}{\textbf{SF 12}}  \\
\cline{4-15}
& & &\multicolumn{1}{|c|}{\textbf{{\begin{tabular}{@{}c@{}}Transmitted \\ Packets\end{tabular}}}} & \multicolumn{1}{|c|}{\textbf{{\begin{tabular}{@{}c@{}}Received \\ Packets\end{tabular}}}} & \multicolumn{1}{|c|}{\textbf{{\begin{tabular}{@{}c@{}}Delivery \\ Ratio (\%)\end{tabular}}}}&\multicolumn{1}{|c|}{\textbf{{\begin{tabular}{@{}c@{}}Transmitted \\ Packets\end{tabular}}}} & \multicolumn{1}{|c|}{\textbf{{\begin{tabular}{@{}c@{}}Received \\ Packets\end{tabular}}}} & \multicolumn{1}{|c|}{\textbf{{\begin{tabular}{@{}c@{}}Delivery \\ Ratio (\%)\end{tabular}}}}&\multicolumn{1}{|c|}{\textbf{{\begin{tabular}{@{}c@{}}Transmitted \\ Packets\end{tabular}}}} & \multicolumn{1}{|c|}{\textbf{{\begin{tabular}{@{}c@{}}Received \\ Packets\end{tabular}}}} & \multicolumn{1}{|c|}{\textbf{{\begin{tabular}{@{}c@{}}Delivery \\ Ratio (\%)\end{tabular}}}}&\multicolumn{1}{|c|}{\textbf{{\begin{tabular}{@{}c@{}}Transmitted \\ Packets\end{tabular}}}} & \multicolumn{1}{|c|}{\textbf{{\begin{tabular}{@{}c@{}}Received \\ Packets\end{tabular}}}} & \multicolumn{1}{|c|}{\textbf{{\begin{tabular}{@{}c@{}}Delivery \\ Ratio (\%)\end{tabular}}}} \\

\hline
\hline
& & & & & & & & & & & & & &\\[0.2ex]
\multicolumn{1}{|c|}{\multirow{5}{*}{LoS Scenario}} & \multicolumn{1}{|c|}{4 km/h}& \multicolumn{1}{|c|}{\begin{tabular}{@{}c@{}} Walking \\ Bremen, Germany\end{tabular}}& \multicolumn{1}{|c|}{749}& \multicolumn{1}{|c|}{690}& \multicolumn{1}{|c|}{\textbf{92.12}} & \multicolumn{1}{|c|}{447}& \multicolumn{1}{|c|}{447}& \multicolumn{1}{|c|}{\textbf{100}}& \multicolumn{1}{|c|}{747}& \multicolumn{1}{|c|}{743}& \multicolumn{1}{|c|}{\textbf{99.46}} & \multicolumn{1}{|c|}{465}& \multicolumn{1}{|c|}{465}& \multicolumn{1}{|c|}{\textbf{100}}\\
\cline{2-15}
& & & & & & & & & & & & & &\\[0.2ex]
 & \multicolumn{1}{|c|}{10 km/h}& \multicolumn{1}{|c|}{\begin{tabular}{@{}c@{}} Bicycle \\ Bremen, Germany\end{tabular}}& \multicolumn{1}{|c|}{705}& \multicolumn{1}{|c|}{650}& \multicolumn{1}{|c|}{\textbf{92.19}} & \multicolumn{1}{|c|}{461}& \multicolumn{1}{|c|}{460}& \multicolumn{1}{|c|}{\textbf{99.78}}& \multicolumn{1}{|c|}{741}& \multicolumn{1}{|c|}{737}& \multicolumn{1}{|c|}{\textbf{99.46}} & \multicolumn{1}{|c|}{469}& \multicolumn{1}{|c|}{469}& \multicolumn{1}{|c|}{\textbf{100}}\\
\cline{2-15}
& & & & & & & & & & & & & &\\[0.2ex]
 & \multicolumn{1}{|c|}{20 km/h}& \multicolumn{1}{|c|}{\begin{tabular}{@{}c@{}} E-Scooter \\ Bremen, Germany\end{tabular}}& \multicolumn{1}{|c|}{788}& \multicolumn{1}{|c|}{722}& \multicolumn{1}{|c|}{\textbf{91.62}} & \multicolumn{1}{|c|}{472}& \multicolumn{1}{|c|}{472}& \multicolumn{1}{|c|}{\textbf{100}}& \multicolumn{1}{|c|}{736}& \multicolumn{1}{|c|}{723}& \multicolumn{1}{|c|}{\textbf{98.23}} & \multicolumn{1}{|c|}{462}& \multicolumn{1}{|c|}{456}& \multicolumn{1}{|c|}{\textbf{98.70}}\\
\cline{2-15}
& & & & & & & & & & & & & &\\[0.2ex]
 & \multicolumn{1}{|c|}{40 km/h}& \multicolumn{1}{|c|}{\begin{tabular}{@{}c@{}} Drone \\ Colombo, Sri Lanka\end{tabular}}& \multicolumn{1}{|c|}{589}& \multicolumn{1}{|c|}{579}& \multicolumn{1}{|c|}{\textbf{98.30}} & \multicolumn{1}{|c|}{424}& \multicolumn{1}{|c|}{414}& \multicolumn{1}{|c|}{\textbf{97.64}}& \multicolumn{1}{|c|}{567}& \multicolumn{1}{|c|}{555}& \multicolumn{1}{|c|}{\textbf{97.88}} & \multicolumn{1}{|c|}{265}& \multicolumn{1}{|c|}{256}& \multicolumn{1}{|c|}{\textbf{96.60}}\\
\cline{2-15}
& & & & & & & & & & & & & &\\[0.2ex]
 & \multicolumn{1}{|c|}{58 km/h}& \multicolumn{1}{|c|}{\begin{tabular}{@{}c@{}} Drone \\ Colombo, Sri Lanka\end{tabular}}& \multicolumn{1}{|c|}{575}& \multicolumn{1}{|c|}{561}& \multicolumn{1}{|c|}{\textbf{97.56}} & \multicolumn{1}{|c|}{314}& \multicolumn{1}{|c|}{309}& \multicolumn{1}{|c|}{\textbf{98.40}}& \multicolumn{1}{|c|}{481}& \multicolumn{1}{|c|}{430}& \multicolumn{1}{|c|}{\textbf{89.39}} & \multicolumn{1}{|c|}{136}& \multicolumn{1}{|c|}{132}& \multicolumn{1}{|c|}{\textbf{97.05}}\\
\hline 
\hline
& & & & & & & & & & & & & &\\[0.2ex]
\multicolumn{1}{|c|}{\multirow{3}{*}{LoS + Non-LoS Scenario}} & \multicolumn{1}{|c|}{4 km/h}& \multicolumn{1}{|c|}{\begin{tabular}{@{}c@{}} Walking \\ Bremen, Germany\end{tabular}}& \multicolumn{1}{|c|}{757}& \multicolumn{1}{|c|}{429}& \multicolumn{1}{|c|}{\textbf{56.57}} & \multicolumn{1}{|c|}{452}& \multicolumn{1}{|c|}{392}& \multicolumn{1}{|c|}{\textbf{86.72}}& \multicolumn{1}{|c|}{762}& \multicolumn{1}{|c|}{649}& \multicolumn{1}{|c|}{\textbf{85.17}} & \multicolumn{1}{|c|}{471}& \multicolumn{1}{|c|}{465}& \multicolumn{1}{|c|}{\textbf{98.72}}\\
\cline{2-15}
& & & & & & & & & & & & & &\\[0.2ex]
 & \multicolumn{1}{|c|}{10 km/h}& \multicolumn{1}{|c|}{\begin{tabular}{@{}c@{}} Bicycle \\ Bremen, Germany\end{tabular}}& \multicolumn{1}{|c|}{694}& \multicolumn{1}{|c|}{428}& \multicolumn{1}{|c|}{\textbf{61.67}} & \multicolumn{1}{|c|}{458}& \multicolumn{1}{|c|}{414}& \multicolumn{1}{|c|}{\textbf{90.39}}& \multicolumn{1}{|c|}{754}& \multicolumn{1}{|c|}{676}& \multicolumn{1}{|c|}{\textbf{89.65}} & \multicolumn{1}{|c|}{456}& \multicolumn{1}{|c|}{449}& \multicolumn{1}{|c|}{\textbf{98.46}}\\
\cline{2-15}
& & & & & & & & & & & & & &\\[0.2ex]
 & \multicolumn{1}{|c|}{20 km/h}& \multicolumn{1}{|c|}{\begin{tabular}{@{}c@{}} E-Scooter \\ Bremen, Germany\end{tabular}}& \multicolumn{1}{|c|}{729}& \multicolumn{1}{|c|}{449}& \multicolumn{1}{|c|}{\textbf{61.59}} & \multicolumn{1}{|c|}{461}& \multicolumn{1}{|c|}{458}& \multicolumn{1}{|c|}{\textbf{99.34}}& \multicolumn{1}{|c|}{763}& \multicolumn{1}{|c|}{659}& \multicolumn{1}{|c|}{\textbf{86.36}} & \multicolumn{1}{|c|}{483}& \multicolumn{1}{|c|}{478}& \multicolumn{1}{|c|}{\textbf{98.96}}\\
\cline{2-15}

\hline
\end{tabular}%
}
\end{center}
\caption{Results of the Summer-Experiments}
\vspace{-5mm}
\label{tab:summer}
\end{table*}

\begin{figure}[]
\centering
\begin{subfigure}{0.49\textwidth}
    \includegraphics[width=\columnwidth]{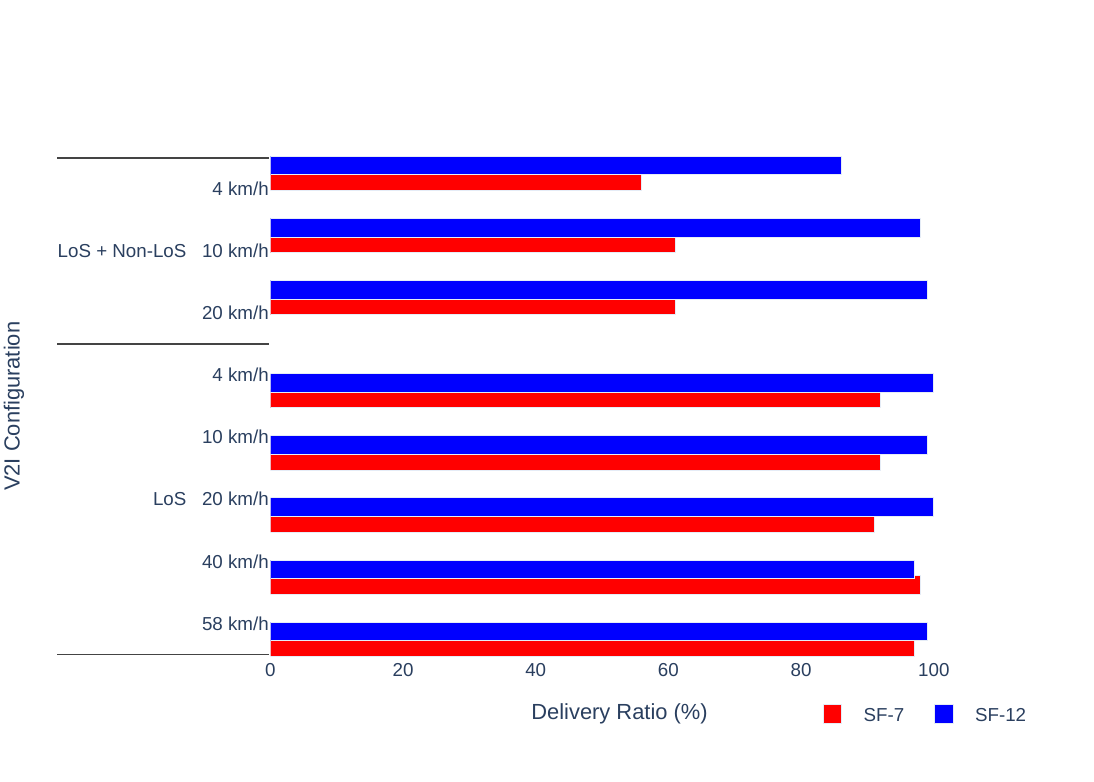}
    \caption{Summer experiments with V2I configuration}
    \label{fig:Summer-first}
\end{subfigure}
\begin{subfigure}{0.49\textwidth}
    \includegraphics[width=\columnwidth]{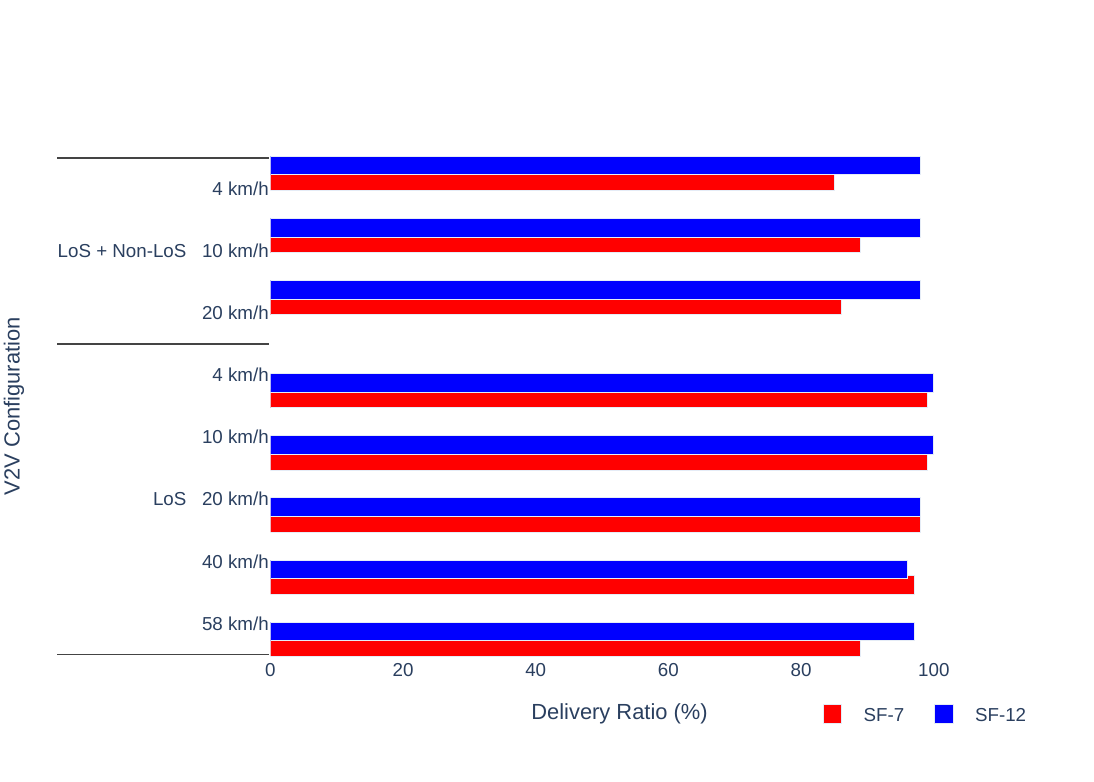}
    \caption{Summer experiments with V2V configuration}
    \label{fig:Summer-second}
\end{subfigure}
        
\caption{Delivery ratios for the Summer experiments}
\label{Summer_V2I_V2V}
\end{figure}

The Summer experiments contain both 'LoS Scenario' and 'LoS + Non-LoS Scenario' as discussed in Section \ref{sec:Experiment Setup}. The obtained results are depicted in Table \ref{tab:summer} and the same results are illustrated graphically in Figure \ref{Summer_V2I_V2V}. We measured the delivery ratio as the ratio between the number of transmitted and received packets. All of our experiments were limited to 30 minutes except the experiments with the drones, which were limited to 15 minutes. However, the number of transmitted packets varies in different experiments because it depends on the hardware behavior, like the time required to obtain the GPS coordinates. Furthermore, the number of packets transmitted with SF-12 is significantly lower compared to SF-7 because SF-12 requires a longer transmission time due to the longer symbol time.

The results with SF-12 have a better delivery ratio than SF-7 in all experiments with both V2I and V2V configurations, as shown in Figure \ref{fig:Summer-first} and Figure \ref{fig:Summer-second}. Furthermore, the improvement with SF-12 compared to SF-7 is more significant in the 'LoS + Non-LoS Scenario'. This improvement is due to the longer symbol times of SF-12, which enable the detecting of weaker signals more than SF-7. In addition, the wider bandwidth of SF-12 helps to reduce interference from other signals and noise, especially in crowded environments. However, SF-12 has the trade-off of a low data rate and, therefore, may not be suitable for some VANET applications. During our experiments, we used the frequency of 868 MHz for LoRa communication. However, in Europe, LoRa communication is also allowed to work on the frequency of 433 MHz. Since the lower frequencies are better at penetrating obstacles, the delivery ratios can be expected to increase slightly with 433 MHz.

Furthermore, the delivery ratios for the 'LoS + Non-LoS scenario' was improved with V2V configuration (Figure \ref{fig:Summer-second}) than with V2I configuration (Figure \ref{fig:Summer-first}). This improvement is caused by our experiment scenario because, as discussed in Section \ref{sec:Experiment Setup} in the V2V configuration, we moved the devices on the same route in opposite directions. Therefore, transmitter and receiver meet multiple times during one experiment depending on their relative speed, resulting in better delivery ratios than the V2I configuration.

\begin{figure}[]
\centering
\begin{subfigure}{0.5\textwidth}
    \includegraphics[width=\textwidth]{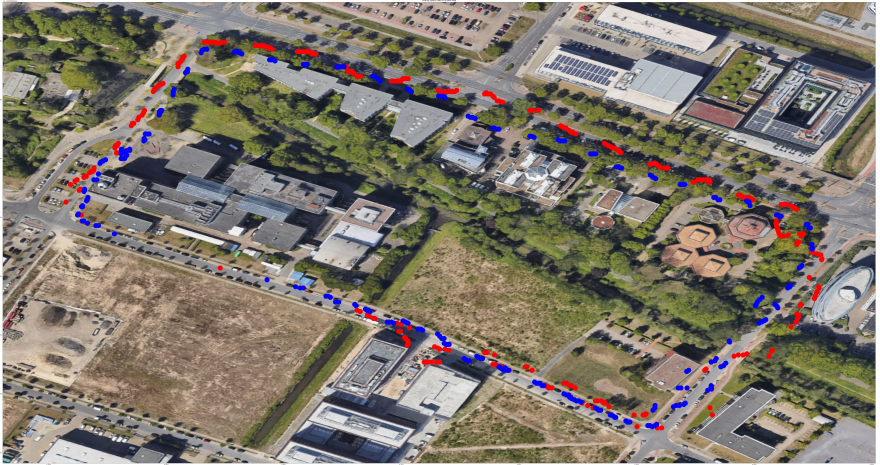}
    \caption{SF-7 and SF-12 with 4 km/h}
    \label{fig:first_Nw1}
\end{subfigure}
\begin{subfigure}{0.5\textwidth}
    \includegraphics[width=\textwidth]{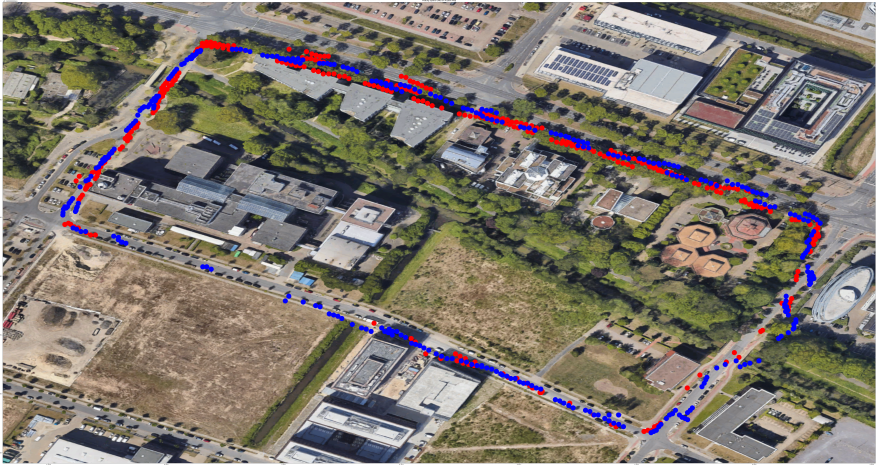}
    \caption{SF-7 and SF-12 with 10 km/h}
    \label{fig:second_Nw1}
\end{subfigure}
\begin{subfigure}{0.5\textwidth}
    \includegraphics[width=\textwidth]{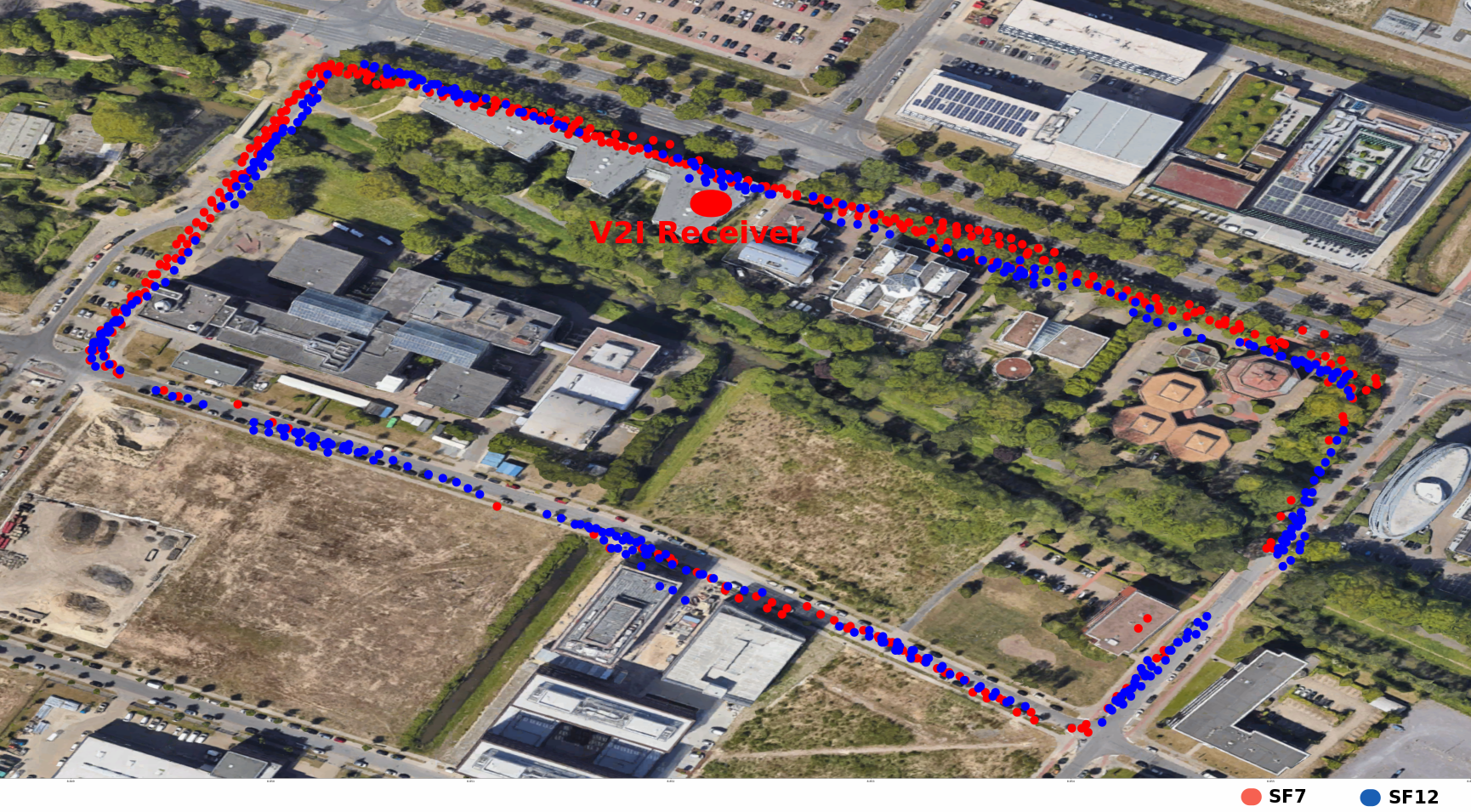}
    \caption{SF-7 and SF-12 with 20 km/h}
    \label{fig:third_Nw1}
\end{subfigure}
        
\caption{The distribution of received packets in 'LoS + Non-LoS scenario' with V2I configuration}
\label{NW1_Summer_V21_4_10_20}
\end{figure}

To better visualise the effect of LoS condition on LoRa with SF-7 and SF-12, we mapped the distribution of received packets for the 'LoS + Non-LoS scenario' in the V2I configuration in Figure \ref{NW1_Summer_V21_4_10_20}. The mapped data indicate that with the SF-7, most of the packets were not received when the signal was attenuated by surrounding objects (mainly the University building) between the transmitter and the receiver. This behavior is shown in Figure \ref{NW1_Summer_V21_4_10_20} with very few red dots in the area around the University building that obstruct the communication. In contrast, in the same area, there are many more blue dots, which implies that SF-12 manages to transmit packets successfully. Furthermore, there are more blue dots around the University building with 20 km/h (Figure \ref{fig:third_Nw1}) than 10 km/h (Figure \ref{fig:second_Nw1}) and 4 km/h (Figure \ref{fig:first_Nw1}) because with higher speeds the transmitter was able to travel multiple rounds on the route allowing more transmissions to be successful around the University building.

In all the conducted experiments LoRa communication was tested with speeds up to 58 km/h. In these considered speeds, the delivery ratios of LoRa communication were not affected by the induced Doppler shift, as shown in Table \ref{tab:summer} and Figure \ref{Summer_V2I_V2V}. This implies that LoRa communication is robust against the urban VANET speeds. However, our results show that LoS affects the performance of LoRa significantly, and SF-12 performs better in the case of Non-LoS communication than SF-7.

\subsection{Fall experiments}
\label{subsec:Fall-Experiments}

As explained in Section \ref{sec:Experiment Setup}, our experiments with the LoS scenario consist of several trees along the route, which might slightly affect the LoS condition. In order to evaluate the effect of these trees on LoRa communication, we extended our 'LoS Scenario' by repeating the same set of Summer experiments in the following Fall season.  

\begin{table*}
\begin{center}
\resizebox{\textwidth}{!}{
\begin{tabular}{|l|l|l|l|l|l|l|l|l|l|l|l|l|l|}
\hline
\multicolumn{1}{|c|}{\multirow{2}{*}{\textbf{Scenario}}}& \multicolumn{1}{|c|}{\multirow{2}{*}{\textbf{Speed}}}& \multicolumn{6}{|c|}{\textbf{LoS-Summer}} & \multicolumn{6}{|c|}{\textbf{LoS-Fall}}  \\
\cline{3-14}
& & \multicolumn{3}{|c|}{\textbf{SF 7}} & \multicolumn{3}{|c|}{\textbf{SF 12}}& \multicolumn{3}{|c|}{\textbf{SF 7}} & \multicolumn{3}{|c|}{\textbf{SF 12}}  \\
\cline{3-14}
& &\multicolumn{1}{|c|}{\textbf{{\begin{tabular}{@{}c@{}}Transmitted \\ Packets\end{tabular}}}} & \multicolumn{1}{|c|}{\textbf{{\begin{tabular}{@{}c@{}}Received \\ Packets\end{tabular}}}} & \multicolumn{1}{|c|}{\textbf{{\begin{tabular}{@{}c@{}}Delivery \\ Ratio (\%)\end{tabular}}}}&\multicolumn{1}{|c|}{\textbf{{\begin{tabular}{@{}c@{}}Transmitted \\ Packets\end{tabular}}}} & \multicolumn{1}{|c|}{\textbf{{\begin{tabular}{@{}c@{}}Received \\ Packets\end{tabular}}}} & \multicolumn{1}{|c|}{\textbf{{\begin{tabular}{@{}c@{}}Delivery \\ Ratio (\%)\end{tabular}}}}&\multicolumn{1}{|c|}{\textbf{{\begin{tabular}{@{}c@{}}Transmitted \\ Packets\end{tabular}}}} & \multicolumn{1}{|c|}{\textbf{{\begin{tabular}{@{}c@{}}Received \\ Packets\end{tabular}}}} & \multicolumn{1}{|c|}{\textbf{{\begin{tabular}{@{}c@{}}Delivery \\ Ratio (\%)\end{tabular}}}}&\multicolumn{1}{|c|}{\textbf{{\begin{tabular}{@{}c@{}}Transmitted \\ Packets\end{tabular}}}} & \multicolumn{1}{|c|}{\textbf{{\begin{tabular}{@{}c@{}}Received \\ Packets\end{tabular}}}} & \multicolumn{1}{|c|}{\textbf{{\begin{tabular}{@{}c@{}}Delivery \\ Ratio (\%)\end{tabular}}}} \\
\hline
\hline
& & & & & & & & & & & & & \\[0.2ex]
\multicolumn{1}{|c|}{\multirow{3}{*}{V2I Configuration}} & \multicolumn{1}{|c|}{4 km/h}& \multicolumn{1}{|c|}{749}& \multicolumn{1}{|c|}{690}& \multicolumn{1}{|c|}{\textbf{92.12}} & \multicolumn{1}{|c|}{447}& \multicolumn{1}{|c|}{447}& \multicolumn{1}{|c|}{\textbf{100}}& \multicolumn{1}{|c|}{833}& \multicolumn{1}{|c|}{813}& \multicolumn{1}{|c|}{\textbf{97.59}} & \multicolumn{1}{|c|}{457}& \multicolumn{1}{|c|}{457}& \multicolumn{1}{|c|}{\textbf{100}}\\
\cline{2-14}
& & & & & & & & & & & & & \\[0.2ex]
& \multicolumn{1}{|c|}{10 km/h} & \multicolumn{1}{|c|}{705}& \multicolumn{1}{|c|}{650}& \multicolumn{1}{|c|}{\textbf{92.19}} & \multicolumn{1}{|c|}{461}& \multicolumn{1}{|c|}{460}& \multicolumn{1}{|c|}{\textbf{99.78}}& \multicolumn{1}{|c|}{764}& \multicolumn{1}{|c|}{752}& \multicolumn{1}{|c|}{\textbf{98.42}} & \multicolumn{1}{|c|}{454}& \multicolumn{1}{|c|}{454}& \multicolumn{1}{|c|}{\textbf{100}}\\
\cline{2-14}
& & & & & & & & & & & & & \\[0.2ex]
& \multicolumn{1}{|c|}{20 km/h} & \multicolumn{1}{|c|}{788}& \multicolumn{1}{|c|}{722}& \multicolumn{1}{|c|}{\textbf{91.62}} & \multicolumn{1}{|c|}{472}& \multicolumn{1}{|c|}{472}& \multicolumn{1}{|c|}{\textbf{100}}& \multicolumn{1}{|c|}{753}& \multicolumn{1}{|c|}{744}& \multicolumn{1}{|c|}{\textbf{98.80}} & \multicolumn{1}{|c|}{522}& \multicolumn{1}{|c|}{522}& \multicolumn{1}{|c|}{\textbf{100}}\\
\hline
\hline
& & & & & & & & & & & & & \\[0.2ex]
\multicolumn{1}{|c|}{\multirow{3}{*}{V2V Configuration}} & \multicolumn{1}{|c|}{4 km/h}& \multicolumn{1}{|c|}{747}& \multicolumn{1}{|c|}{743}& \multicolumn{1}{|c|}{\textbf{99.46}} & \multicolumn{1}{|c|}{465}& \multicolumn{1}{|c|}{465}& \multicolumn{1}{|c|}{\textbf{100}}& \multicolumn{1}{|c|}{833}& \multicolumn{1}{|c|}{822}& \multicolumn{1}{|c|}{\textbf{98.67}} & \multicolumn{1}{|c|}{457}& \multicolumn{1}{|c|}{456}& \multicolumn{1}{|c|}{\textbf{99.78}}\\
\cline{2-14}
& & & & & & & & & & & & & \\[0.2ex]
& \multicolumn{1}{|c|}{10 km/h} & \multicolumn{1}{|c|}{741}& \multicolumn{1}{|c|}{737}& \multicolumn{1}{|c|}{\textbf{99.46}} & \multicolumn{1}{|c|}{469}& \multicolumn{1}{|c|}{469}& \multicolumn{1}{|c|}{\textbf{100}}& \multicolumn{1}{|c|}{764}& \multicolumn{1}{|c|}{751}& \multicolumn{1}{|c|}{\textbf{98.29}} & \multicolumn{1}{|c|}{454}& \multicolumn{1}{|c|}{453}& \multicolumn{1}{|c|}{\textbf{99.77}}\\
\cline{2-14}
& & & & & & & & & & & & & \\[0.2ex]
& \multicolumn{1}{|c|}{20 km/h} & \multicolumn{1}{|c|}{736}& \multicolumn{1}{|c|}{723}& \multicolumn{1}{|c|}{\textbf{98.23}} & \multicolumn{1}{|c|}{462}& \multicolumn{1}{|c|}{456}& \multicolumn{1}{|c|}{\textbf{98.70}}& \multicolumn{1}{|c|}{898}& \multicolumn{1}{|c|}{897}& \multicolumn{1}{|c|}{\textbf{99.88}} & \multicolumn{1}{|c|}{522}& \multicolumn{1}{|c|}{522}& \multicolumn{1}{|c|}{\textbf{100}}\\
\hline

\end{tabular}%
}
\end{center}
\caption{Results of the Fall-Experiments}
\vspace{-5mm}
\label{tab:summer_vs_fall}
\end{table*}

\begin{figure}[]
\centering
\begin{subfigure}{0.49\textwidth}
    \includegraphics[width=\textwidth]{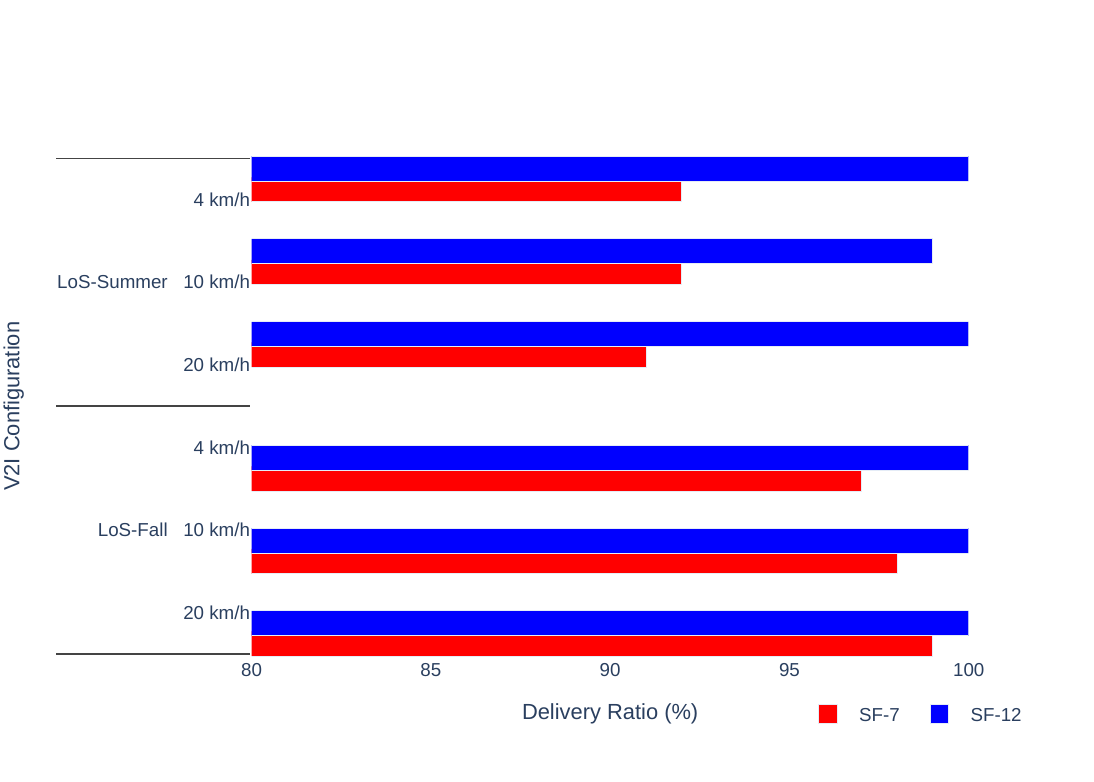}
    \caption{Fall and Summer results with V2I configuration}
    \label{fig:first_summer_vs_fall}
\end{subfigure}
\begin{subfigure}{0.49\textwidth}
    \includegraphics[width=\textwidth]{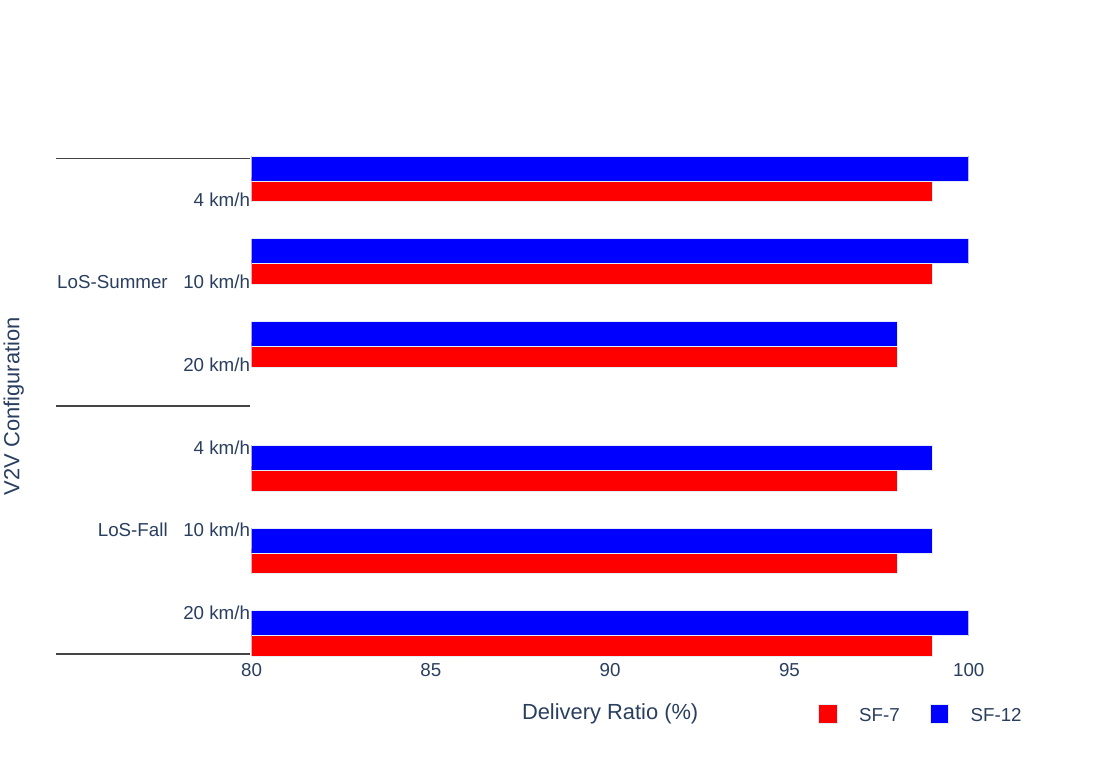}
    \caption{Fall and Summer results with V2V configuration}
    \label{fig:second_summer_vs_fall}
\end{subfigure}
        
\caption{Delivery ratios for the Fall and Summer Experiments}
\label{Los_Summer_vs_Fall}
\end{figure}

Table \ref{tab:summer_vs_fall} summarises these results along with LoS-Summer experiments for comparison, and Figure \ref{Los_Summer_vs_Fall} illustrates the same results graphically. The delivery ratios with SF-7 were noticeably increased with V2I configuration compared to LoS-Summer, as shown in Figure \ref{fig:first_summer_vs_fall}. This improvement is caused by the better LoS condition in the Fall, resulting in less attenuation of the signals. There is no significant difference in results for SF-12 because SF-12 is more robust against attenuated signals, and as expected, SF-12 outperforms SF-7 in all the experiments. Furthermore, as discussed in Section \ref{subsec:Summer-Experiments}, V2V configuration depicts slightly better results than V2I configuration because of the mobility pattern of the devices.

\begin{figure*}[]
\centering
\includegraphics[width=0.8\textwidth]{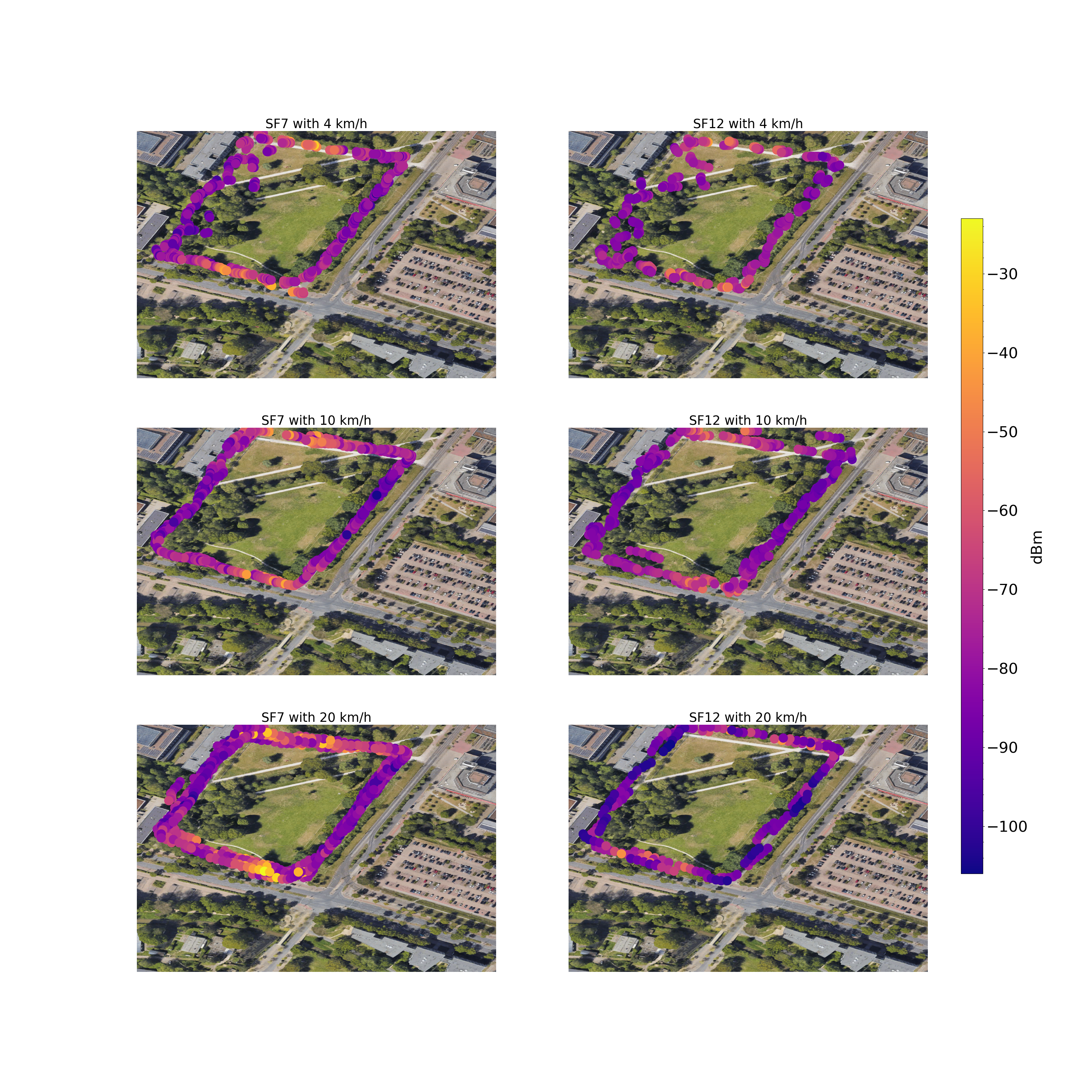}
\caption{The RSSI distribution for LoS-Fall experiments with V2V configuration}
\label{fall_summer_RRSI_V2V}
\end{figure*}

\begin{figure}[]
\centering
\includegraphics[width=0.5\textwidth]{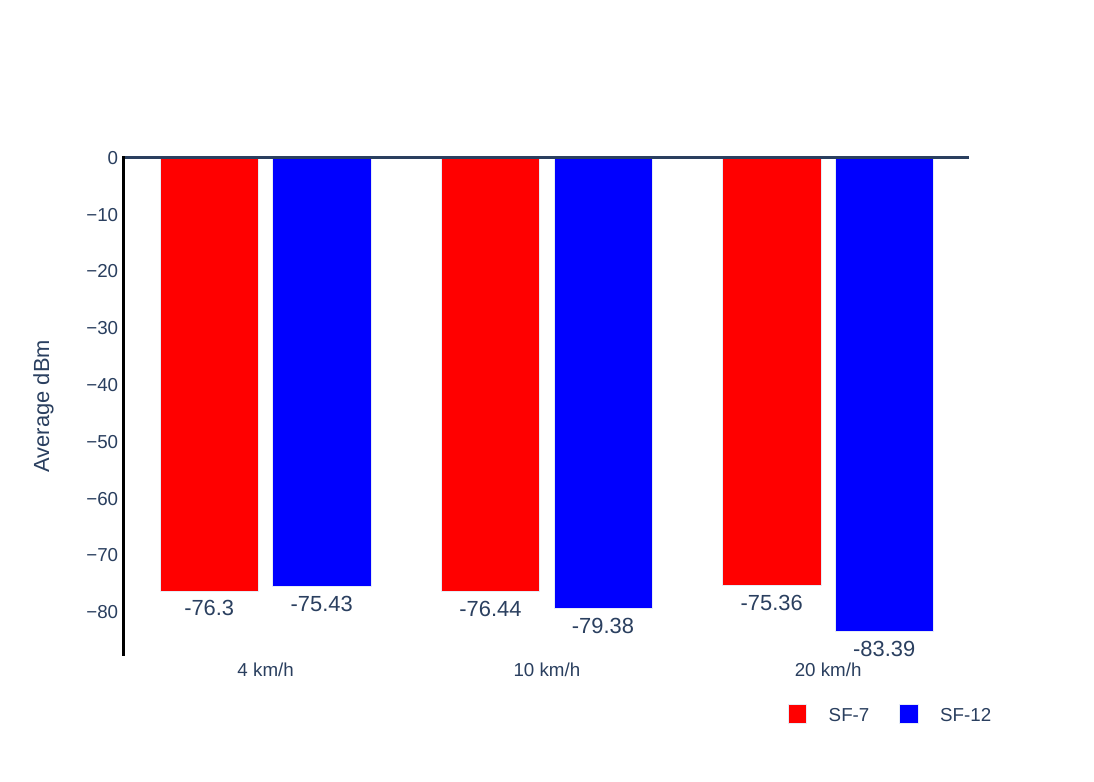}
\caption{Average RSSI for V2V configuration for Fall experiments}
\label{RSSI}
\end{figure}

Moreover, we decided to collect the RSSI values of the received packets for the Fall experiments. The distribution of RSSI values obtained with V2V configuration is illustrated in Figure \ref{fall_summer_RRSI_V2V} and summarised graphically in Figure \ref{RSSI}. Overall, the RSSI values with SF-7 are better than SF-12. These better RSSI values are caused by the shorter transmission time of SF-7, giving less opportunity for the signal to be attenuated during the transmission, and once attenuated, the signal is not received at the receiver. In contrast, SF-12 was able to receive weaker signals than SF-7, as shown in Figure \ref{RSSI}. In addition, the RSSI with SF-12 was more affected by the higher moving speed, as shown in \ref{RSSI} with decreased RSSI values with increasing speed. This implies that the longer transmission time of SF-12 increases the likelihood of frequency shifts and makes it more vulnerable to Doppler shift.

\section{Conclusion}
\label{sec:Section 5}

LoRa communication has gained a lot of attention in the domain of WSNs due to its characteristics like long range and low power. In this study, we investigate the possibility of using LoRa for VANETs in non-safety related applications as a supporting communication technology to ease the burden on the main communication technologies: DSRC and C-V2X. Our experimental analysis was focused on testing LoRa in urban VANET scenarios where the most congested and demanding communication channels are expected. We conducted an extensive set of real-world experiments and tested the LoRa communication behavior in terms of delivery ratio and RSSI. 

Our results show that LoRa communication performs well with urban speeds up to 58 km/h in both V2I and V2V configurations. Therefore, LoRa communication was robust against the Doppler shifts caused by these urban speeds. However, the performance of LoRa depends significantly on LoS communication, and SF-12 depicted better performance than SF-7 in the presence of Non-LoS communication due to its ability to detect weaker signals. Even though SF-12 depicts better delivery ratios, it also showed more vulnerability toward Dopper shifts, resulting in lower RSSI values at higher speeds.

Our results proved that LoRa can be used in urban VANET scenarios for non-safety related applications. However, the most suitable LoRa parameters must be selected depending on the application requirements, such as the required data rate and surrounding environment.

\section{Acknowledgement}
\label{sec:Section 6}

We would like to thank Suleco (Pvt) Ltd and LHP Group for providing the infrastructure for the drone experiments. Furthermore, we thank all the following individuals for providing their support for conducting these multiple experiments. Ajmal Azeez (drone pilot), Asanga Udugama, Jens Dede, Saurabh Band, Cyrille Sepele, Purna Rathnayake, Jehan Jayawardana, and Dinithi Amarawardana.

\bibliographystyle{unsrt}
\bibliography{paper}

\end{document}